\documentclass[a4paper]{article}
\usepackage{amsmath}
\usepackage{amsfonts}
\usepackage{amssymb}
\usepackage{graphicx}
\newcommand{\ds}{\displaystyle}
\newcommand{\RR}{\mathbb{R}}
\newcommand{\Ord}{\mathcal O}
\newcommand{\beq}{\begin{equation}}
\newcommand{\eeq}{\end{equation}}

\newcommand{\de}{\delta}

\def\eps{{\varepsilon}}
\def\del{{\delta}}

\def\eps{{\epsilon}}
\def\del{{\delta}}

\def\eps{{\epsilon}}
\begin{document}
  \title{A novel route to a Hopf-bifurcation scenario in switched systems with dead zone}
  \author{
P. Kowalczyk \thanks{
School of Computing, Mathematics and Digital Technology, John Dalton Building, Manchester Metropolitan University, Chester Street, Manchester, M1 5GD, U.K.,  Email: {\tt p.kowalczyk@mmu.ac.uk}}
}
\maketitle
\abstract{Planar switched system with dead-zone are analyzed. In particular, we consider the effects of perturbation of the linear control law from purely positional to position-velocity control. This type of perturbation leads to a novel Hopf-like discontinuity induced bifurcation. We show that this bifurcation leads to the creation of small scale limit cycle attractors, which scale as the square root of the bifurcation parameter. We note that a Hopf-like bifurcation analyzed in non-smooth systems (see \cite{FrPoRoTo:98}) is characterized by a linear scaling law. We then investigate numerically a planar switched system with positional feedback law, dead-zone and time delay in the switching function. Using the same parameter values as for the switched system without delay in the switching function, we show numerically a Hopf-like bifurcation scenario which matches qualitatively and quantitatively with the scenario analyzed for the non-delayed system.
}
\\
\vspace{0cm}
\newline
\section{Introduction}
\label{sec:introd}
Dynamical systems which are characterized by switchings between a number of distinct differentiable vector fields, with the switching law that depends on the value of some state variable, are common in engineering applications (e.g. control or mechanical engineering \cite{BaVe:01,diBJoVa:01,BuClDu:95,PoHiOs:95}). Such systems, depending on the context, are termed as hybrid dynamical systems, non-smooth systems, or switched systems. In recent years, much of research effort has been spent on classifying bifurcations, termed as discontinuity induced bifurcations -
DIBs for short, specifically pertaining to systems with switched vector fields, see, for some examples, \cite{Le:00,KuRiGr:03,diBNoOl:08,Koetal:06,diBeBuChaKo:08}. However, unlike in the case of $n$-dimensional differentiable vector fields a complete theory of bifurcation scenarios (e.g. local bifurcations of co-dimension one) in non-smooth systems has, as yet, not been possible. Switched (or hybrid systems) can be seen as a concatenation of differentiable vector fields in a way which is dependent on the class of switched systems under investigations. Hence the structure of phase space in non-smooth systems allows a plethora of different configurations even in the case of low dimensional systems. For this reason, bifurcations in non-smooth systems are treated by considering non-smooth (hybrid systems) with a switching law that has a certain structure and thus the conditions that define a switching law provide restrictions as to what types of bifurcations could occur in a given class of switched (hybrid) systems.

Recently, it has been suggested that the presence of switched or intermittent control induces sway dynamics during quiet standing of humans \cite{Mietal:09,Asa:09,GaLoLaGo:11,GaLoGoLa:14}. One way of verifying this hypothesis is to investigate the dynamics of switched models in the context of human balance as presented in \cite{Mietal:09,Asa:09,KoGletal:11,GaLoGoLa:14,MiInSt:15}
 In particular, the authors in \cite{Mietal:09,Asa:09,KoGletal:11,GaLoGoLa:14,MiInSt:15} consider different types of switched control laws to account for sway patterns observed experimentally, and the dynamics of these models is investigated. In \cite{Asa:09} phase space is divided into different adjacent regions where the control action is either switched on or off. The authors argue that the convergence to upright equilibrium during quiet standing is linked with the neuromuscular system directing the body to follow the stable manifold of the saddle type ``upright'' equilibrium. In \cite{KoGletal:11} a multistability and homoclinic bifurcation scenario in a switched model with dead-zone and delay in the switching function and in the delay in state variables is shown.
 These models are furthered analyzed in \cite{SiKuLi:12} where homoclinic bifurcations, complex bursting dynamics, so-called boundary bifurcations are investigated. Another context of human neuromuscular control where switched models could be applied is in modeling variations of threshold detection due to diabetes or aging. Motivated by the above mentioned applications,
 in the current paper, we consider planar switched systems with dead-zone. In particular, within the dead-zone the equations of motions describe an inverted pendulum model and outside of the dead-zone a position or position-velocity control is applied to the inverted pendulum model system.
 The contribution of the current work is the analysis of a novel type of a Hopf-like bifurcation scenario as yet not analyzed in the literature triggered by a perturbations from position to position-velocity control. We also link this bifurcation with a Hopf-like bifurcation, first time reported in the current work, in the system with dead-zone, purely positional feedback law and delay in the switching decision function.

The rest of the paper is outlined as follows. In Sec.~\ref{sec:sysofint} switched systems with dead-zone that we analyze in the paper are introduced. We then, in Sec.~\ref{sec:lclstc}, analyze a Hopf-bifurcation ensuing in the system due to perturbation of the control vector using asymptotic method. The existence and stability of small scale limit cycles born in the bifurcation are analyzed in Sec.~\ref{sec:smsosc}. We then proceed in Sec.~\ref{sec:numvrf} to verify our analytical results by means of numerical investigations of the system around the bifurcation point. In the following Sec.~\ref{sec:dlsys}, using a system with delay in the switching function we numerically investigate a Hopf-like bifurcation which shows striking similarity to the Hopf-like scenario of the non-delayed system. Finally, Sec.~\ref{sec:concl} concludes the paper.

\section{Systems of interest}
\label{sec:sysofint}
Consider a class of systems given by
\begin{eqnarray}
\label{eq:sys1}
\dot x = A_{I} x \quad{\mbox{for}}\quad |Cx| \leq \phi,\\
\label{eq:sys2}
\dot x = A_{O}x \quad{\mbox{for}}\quad |Cx| > \phi,
\end{eqnarray}
where $A_{I}\in \RR^2\times \RR^2$ is a non-singular matrix with the eigenvalues corresponding to a saddle-node equilibrium point, and $A_{O}\in \RR^2\times \RR^2$ is a non-singular matrix with the eigenvalues corresponding to a stable equilibrium point of the focus type. The product of the state vector $x\in \RR^2$ and the constant control row vector $C\in \RR^2$ determines the switching between the two linear vector fields for some fixed and positive value of $\phi$. In what follows, we consider a novel Hopf bifurcation scenario in the above class of systems under the variation of the control vector from $C = C_0$ to $C = C_0^{\eps}$.
\subsection{Planar switched systems with dead-zone}
Consider switched systems where the bifurcation parameter, say $\beta$, is increased from $0$ and implies a change of the control vector $C_0$ from
$C_0 = [-1\quad 0]$ to $C_0^{\eps} = [-1\quad\beta]$, where $\beta = \Ord(\eps)$. This
variation is a change from purely positional feedback control law to
position-velocity feedback law.  Matrices $A_I$, $A_O$, the state vector ${\bf{x}}$ and the width of the dead-zone are given by
\beq
A_I = \left(\begin{array}{cc} 0 & 1 \\ A & 0\end{array}\right),\quad
A_O = \left(\begin{array}{cc} 0 & 1 \\ A-K_p & -K_d\end{array}\right),\quad {\bf{x}} =
\left(\begin{array}{c} \theta\\\dot\theta\end{array}\right),\quad |{C_0\bf{x}}|\leq \theta^*,\,\,\mbox{or}\,\,
|{C_0^\eps\bf{x}}|\leq \theta^*,
\label{eq:gensw}
\eeq
where $K_p > A > 0$, $K_p - A > K_d^2/4$, $K_d > 0$ and $\theta^* > 0$. In this set up the eigenvalues of $A_I$ correspond to the system's equilibrium point of a saddle type and the
eigenvalues of $A_O$ correspond to a stable focus, as assumed earlier. Matrix $A_I$ is expressed in so-called controllable canonical form \cite{BaCa:85,Br:91} and this structure can be assumed without loss of generality
\subsection{Phase space topology}
We now define switching manifolds $\Sigma_\pm$ and
$\Sigma_\pm^\eps$ as
\begin{eqnarray}
\label{eq:pC1}
& &\Sigma_\pm =\{(\theta,\,\dot\theta)\in\RR^2\, : \, \pm\theta^* - \theta = 0  \},\\
\label{eq:pC2}
& &\Sigma_\pm^\eps =\{(\theta,\,\dot\theta)\in\RR^2\, : \, \beta\dot\theta \pm \theta^* - \theta = 0  \}
\end{eqnarray}
and regions
\begin{eqnarray}
& & G_{IN} = \{(\theta,\,\dot\theta)\in\RR^2\, : |\theta| < \theta^*\},\\
& & G_{OUT} = \{(\theta,\,\dot\theta)\in\RR^2\, : |\theta| > \theta^*\},\\
& & G_{IN}^\eps = \{(\theta,\,\dot\theta)\in\RR^2\, : |\beta\dot\theta - \theta| < \theta^*\},\\
& & G_{OUT}^\eps = \{(\theta,\,\dot\theta)\in\RR^2\, : |\beta\dot\theta - \theta| > \theta^*\}.
\end{eqnarray}
The flow within region $G_{IN}$ or $G_{IN}^\eps$, say $\psi_{IN}$, is given by the solution of the differential equation
\beq
\ddot\theta - A\theta = 0,
\label{eq:IF}
\eeq
and the flow within region $G_{OUT}$ or $G_{OUT}^\eps$, say $\psi_{OUT}$, is given by the solution of
\beq
\ddot\theta + K_d\dot\theta + (K_p - A)\theta = 0.
\label{eq:OF}
\eeq

Finally, define
\beq
H(\theta,\dot\theta) = \beta\dot\theta + \theta^* - \theta.
\label{eq:swS}
\eeq
Clearly, the zero level set of $H$ defines the switching manifold $\Sigma_+^\eps$.
\section{Local stability calculations}
\label{sec:lclstc}
 It has been shown in \cite{KoGletal:11} that the system dynamics with the switching law given by $C_0{\bf x} = \pm\theta^*$ is governed by the existence of a pair of stable pseudo-equilibria $EQ_\pm  = (\pm\theta^*,\, 0)$. We define a pseudo-equilibrium point in the switched system (\ref{eq:sys1}) and (\ref{eq:sys2}) as a point $EQ\in\Sigma_\pm$ such that for some $t_0$ $(\pm\theta^*(t_0),\, 0)\in\Sigma_\pm$ and  $\forall t \geq t_0$ $(\pm\theta(t),\, 0) = (\pm\theta^*(t_0),\, 0)$. We note that considering only forward time suffices for our purposes. The pseudo-equilibria $(\pm\theta^*(t_0),\, 0)\in\Sigma_\pm$ have been shown to be the only two global attractors of the system.
If we now ``switch on'' the control vector $C_0^\eps$ the two pseudo-equilibria loose their stability and there is born a pair of stable limit cycles with the amplitude which grows in the $\Ord(\sqrt{\beta})$. Hence a Hopf-like bifurcation takes place in the system.
Due to the system's symmetry the Hopf-bifurcation occurs simultaneously around both pseudo-equilibria. It is sufficient to consider the Hopf bifurcation around one
of the two pseudo-equilibria. In what follows we will concentrate on $EQ_+$.
\subsection{Calculating flow time $\tau$}
Consider initial point, say  $P_0 = (\theta(0),\dot\theta(0))\in\Sigma^\eps_+$, such that $\dot\theta(0)= \dot\theta_0 =\Ord(\eps)$.
It then follows that $\theta^*-\theta_0 = -\beta\dot\theta_0=\Ord(\eps^2)$.

If $\theta^*-\theta_0 > 0$ and $\beta > 0$ then $\dot\theta_0 < 0$ and the flow within the dead-zone, in the neighborhood of the pseudo-equilibrium, is governed by the solution
of
$$
\ddot\theta - A\theta = 0,
$$
which, clearly, is given by
\begin{eqnarray}
& & \theta(\tau) = C_1\exp(-\sqrt{A}\tau) + C_2\exp(\sqrt{A}\tau)
\\
& & \dot\theta(\tau) = -\sqrt{A}C_1\exp(-\sqrt{A}\tau) + \sqrt{A}C_2\exp(\sqrt{A}\tau).
\end{eqnarray}
Using the initial conditions we find
\begin{eqnarray}
& & \theta(\tau) = \frac{\sqrt{A}\theta_0-\dot\theta_0}{2\sqrt{A}}\exp(-\sqrt{A}\tau) + \frac{\sqrt{A}\theta_0+\dot\theta_0}{2\sqrt{A}}\exp(\sqrt{A}\tau)
\\
& & \dot\theta(\tau) = -\frac{\sqrt{A}\theta_0-\dot\theta_0}{2}\exp(-\sqrt{A}\tau) + \frac{\sqrt{A}\theta_0+\dot\theta_0}{2}\exp(\sqrt{A}\tau).
\end{eqnarray}

 We assume that the time, say $\tau$, required for the flow to reach $\Sigma_+^\eps$ is $\tau = \Ord(\eps)$. We then get
\begin{eqnarray}
& & \theta(\tau) = \theta_0 - \frac{\theta^* - \theta_0}{\beta}\tau + \frac{1}{2}A\theta_0\tau^2 + \Ord(\eps^3),\\
& & \dot\theta(\tau) = A\theta_0\tau - \frac{\theta^*-\theta_0}{\beta} + \Ord(\eps^2).
\end{eqnarray}
Solving $H = 0$ for $\tau$, to leading order in $\tau$, gives
\beq
\tau = 2\beta + \frac{2(\theta^* - \theta_0)}{A\theta_0\beta} + \Ord(\eps^2).
\eeq

Thus the point of intersection with $\Sigma_+^\eps$, say $P_1\in\Sigma_+^\eps$, is given by
\begin{eqnarray}
\label{eq:psl1}
& & \theta_1 = \theta_0 +  2(\theta^* - \theta_0) + 2A\theta_0\beta^2 > \theta^*, \\
\label{eq:vsl1}
& & \dot\theta_1 = 2A\theta_0\beta + \frac{\theta^* - \theta_0}{\beta} > 0.
\end{eqnarray}
\subsection{Calculating flow time $\delta$}
We now need to find the flow time, say $\delta$, back to the switching line
$\Sigma_+^\eps$ following the $\psi_{OUT}$ flow, which is given by the solution of the differential
equation (\ref{eq:OF})
$$
\ddot\theta + K_d\dot\theta + (K_p  - A)\theta = 0.
$$
Assume $\delta = \Ord(\eps)$ and let $a = K_d/2$ and $b = \sqrt{(K_p - A) - (K_d/2)^2}$.

We find the flow solutions
\begin{eqnarray}
& & \theta(\de) = \theta_1\exp(-a\de)\cos(b\de) + \frac{\dot\theta_1 + a\theta_1}{b}\exp(-a\de)\sin(b\de)
\label{eq:fl2sth}\\
& & \dot\theta(\de) = \dot\theta_1\exp(-a\de)\cos(b\de) - \frac{a\dot\theta_1 + a^2\theta_1 + b^2\theta_1}{b}\exp(-a\de)\sin(b\de).
\label{eq:fl2sdth}
\end{eqnarray}
Expanding to $\Ord(\eps^3)$ equation (\ref{eq:fl2sth}) for the angular position, and equation (\ref{eq:fl2sdth}) for the angular velocity, we get
\begin{eqnarray}
& & \theta(\de) = \theta_1 - \frac{1}{2}(a^2 + b^2)\theta_1\de^2 + \dot\theta_1\de + \Ord(\eps^3)\label{eq:p1}
\\ & & \dot\theta(\de) =  - (a^2 + b^2)\theta_1\de + \dot\theta_1 + \Ord(\eps^2).\label{eq:v1}
\end{eqnarray}
Inserting (\ref{eq:psl1}) and (\ref{eq:vsl1}), for the initial position and velocity components, into (\ref{eq:p1}) and (\ref{eq:v1}) gives
 \begin{eqnarray}
 \label{eq:fl2thb}
& &  \theta(\de) = \theta_0 +  2(\theta^* - \theta_0) + 2A\theta_0\beta^2 - \frac{1}{2}(a^2 + b^2)\theta_0\de^2 + \\
& &  2A\theta_0\beta\de + \frac{\theta^* - \theta_0}{\beta}\de + \Ord(\eps^3)\nonumber\\
\label{eq:fl2dthb}
& & \dot\theta(\del) = - (a^2 + b^2)\theta_0\del + 2A\theta_0\beta + \frac{\theta^* - \theta_0}{\beta} + \Ord(\eps^2).
 \end{eqnarray}
 We may now use the above expressions and insert them in the equation for the switching line (\ref{eq:swS}). We find the time $\delta$
 \beq
 \label{eq:tdel}
 \del = 2\frac{(a^2 + b^2 + 2A)\theta_0\beta^2 + \theta^* - \theta_0}{\beta(a^2 + b^2)\theta_0} + \Ord(\eps^2),
 \eeq
which is well-defined since $a^2 + b^2 = K_p - A > 0$, $\beta = \Ord(\eps) > 0$ and $\theta_0 > 0 = \Ord(1)$.
 Inserting for the time $\delta$ equation (\ref{eq:tdel}) into equations (\ref{eq:fl2thb}) and (\ref{eq:fl2dthb}) for position and velocity, we finally find
 \begin{eqnarray}
 & & \theta(\delta) = \theta_0 - 2\theta_0\beta^2(a^2 + b^2 + A) + \Ord(\eps^3) < \theta_0\\
& & \dot\theta(\delta) = -2\theta_0\beta(a^2 + b^2 + A) - \frac{\theta^* - \theta_0}{\beta} + \Ord(\eps^2) < - \frac{\theta^* - \theta_0}{\beta},
 \end{eqnarray}
 which indicates expansion and hence loss of stability of the pseudo-equilibrium $PS = (\theta^*, 0)$ for small positive $\beta$.
 \section{Small scale stable oscillations}
 \label{sec:smsosc}
 \subsection{Existence of limit cycle}
 To determine the existence and stability of small scale oscillations born in the bifurcation,
 we may use the Hamiltonian function (similarly as in \cite{KoGletal:11})
 \beq
 L(\theta,\dot\theta) = \frac{1}{2}\dot\theta^2 - \frac{1}{2}A\theta^2.
 \eeq
 Consider an initial point, say $P_1(\theta_1,\dot\theta_1)\in\Sigma_+^\eps$, such that $\dot\theta_1  = -\frac{\ds\theta^* - \theta_1}{\ds\beta}> 0$. We seek to find the final point, say $P_2$, such that $\Delta L = L(P_2) - L(P_1) = 0$ and $P_2\in\Sigma_+^\eps$. We call the flow time from $P_1$ to $P_2$ by $\delta$. We assume that the bifurcation has a Hopf character which implies that $(\theta_1,\dot\theta_1)$ are $\Ord(1)$ and $\Ord(\sqrt{\eps})$ respectively. It then follows that the flow time $\delta = \Ord(\sqrt{\eps})$.

 Along any given segment of trajectory generated by flow $\psi_{IN}$ there is no change in the value of Hamiltonian $L$. We use flow $\psi_{OUT}$ and in particular equations (\ref{eq:p1}) and (\ref{eq:v1}) to determine point $P_2$.
 Thus
 \begin{eqnarray}
 \label{eq:DelL1}
 & & \Delta L = \frac{1}{2}(-(a^2 + b^2)\theta_1\de + \dot\theta_1)^2 - A\frac{1}{2}(\theta_1 - \frac{1}{2}(a^2 + b^2)\theta_1\de^2 + ... \\ \nonumber & &  + \dot\theta_1\de)^2 - (\frac{1}{2}\dot\theta_1^2 - \frac{1}{2}A\theta_1^2),
 \end{eqnarray}
 where $(\theta_1,\dot\theta_1)\in\Sigma_+^\eps$.
 We can write $\Delta L$ up to and including terms of $\Ord(\eps)$.
 We have
 \begin{eqnarray}
 \label{eq:DelL2}
 & & \Delta L = \frac{1}{2}((a^2 + b^2)^2\theta_1^2\de^2 + \dot\theta_1^2 - 2(a^2 + b^2)\theta_1\dot\theta_1\de) + ... \\
& &\nonumber - A\frac{1}{2}(\theta_1^2 - \theta_1^2\de^2(a^2 + b^2) + 2\theta_1\dot\theta_1\de) - \frac{1}{2}\dot\theta_1^2 + \frac{1}{2}A\theta_1^2.
 \end{eqnarray}
 Simplifying (\ref{eq:DelL2}) we get
 \beq
\Delta L = -[(a^2 + b^2) + A]\theta_1\dot\theta_1\delta + \frac{1}{2}(a^2 + b^2)\theta_1^2\delta^2[(a^2 + b^2) + A],
 \label{eq:DelL3}
 \eeq
 which after further simplifications leads to
 \beq
 \Delta L = (-\dot\theta_1 + \frac{1}{2}K_p(K_p - A)\theta_1\delta)K_p\theta_1\delta.
 \label{eq:DelL4}
 \eeq
 Using (\ref{eq:DelL3}) we can solve $\Delta L = 0$ for $\delta$, which to leading order in $\eps$ gives
 \beq
 \label{eq:delT}
 \delta = \frac{2\dot\theta_1}{(Kp-A)\theta_1} + \Ord(\eps).
\eeq
We also require $(\theta_1,\dot\theta_1)\in\Sigma_+^\eps$ and $(\theta_2,\dot\theta_2)\in\Sigma_+^\eps$. 
Using equation (\ref{eq:v1}) for the velocity $\dot\theta$ and inserting it into equation for the switching line $\Sigma_+^\eps$ given by (\ref{eq:swS}), we find
\beq
\beta\dot\theta_1 + \theta^* - \theta_1 + \frac{1}{2}(K_p - A)\theta_1\delta^2 - \dot\theta_1\delta + \Ord(\eps^{3/2}) = 0,
\label{eq:swm2}
\eeq
where $\Ord(\eps^{3/2})$ signifies the remaining terms of $\Ord(\eps^{3/2})$ and higher. Note that $\beta\dot\theta_1 + \theta^* - \theta_1 = 0$, but $\beta\dot\theta_1=\Ord(\eps^{3/2})$ and
$\theta^* - \theta_1=\Ord(\eps^{3/2})$. Thus the point $(\theta_2,\dot\theta_2)\in\Sigma_+^\eps$.
\subsection{Stability calculations}
To determine the stability of the limit cycle, we compute
$$
\frac{d\Delta L}{d \dot\theta_1} = \frac{\partial \Delta L}{\partial \dot\theta_1} +
\frac{\partial \Delta L}{\partial\delta}\frac{d \delta}{d\dot\theta_1}.
$$
Using
$$
H(\dot\theta_1,\,\,\delta) = 0
$$
and applying the Implicit Function Theorem, we find
$$
\frac{d\delta}{d \dot\theta_1} = -\frac{\partial H}{\partial \dot\theta_1}\left(\frac{\partial H}{\partial\delta}\right)^{-1}.
$$
Using (\ref{eq:delT}) for $\delta$, we have
\beq
\frac{d\Delta L}{d \dot\theta_1} = -\frac{2K_p\dot\theta_1}{K_p-A} - K_p\theta_1\dot\theta_1\frac{d\delta}{d\dot\theta_1}  + 2K_p\theta_1\dot\theta_1\frac{d\delta}{d\dot\theta_1},
\label{eq:lcs}
\eeq
and so we find
$$
\frac{d\delta}{d\dot\theta_1} = \frac{2}{(K_p - A)\theta_1}
$$
to leading order.

Thus (\ref{eq:lcs}) simplifies to
\beq
\frac{d\Delta L}{d \dot\theta_1} = -\frac{K_p\dot\theta_1}{K_p - A} < 0
\eeq
to leading order, and hence the limit cycle born in the bifurcation is stable.
\begin{figure}
\centering
\includegraphics[scale=0.35]{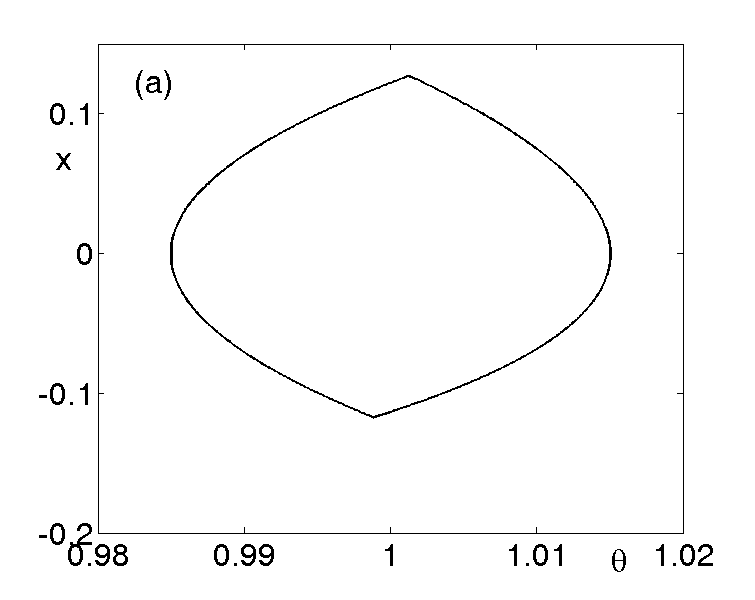}
\includegraphics[scale=0.35]{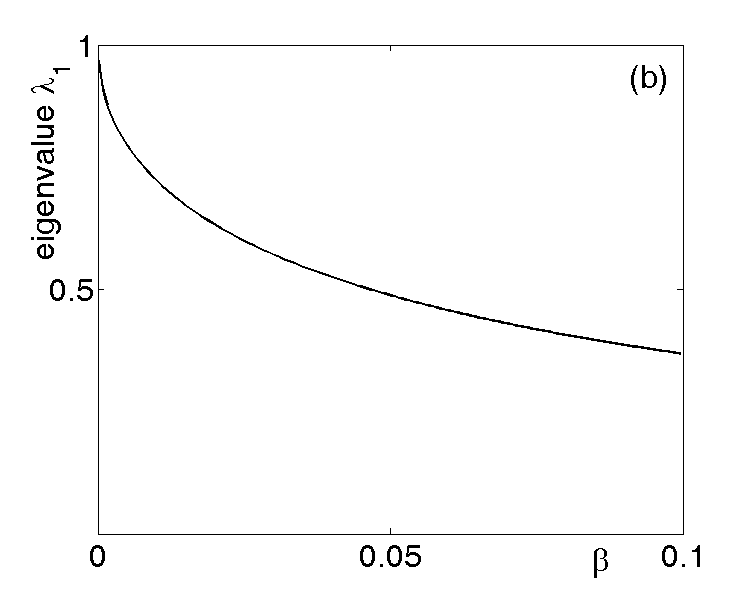}
\caption{(a)A limit cycle attractor. (b)The variation of the non-trivial Floquet multiplier of the limit cycle attractor against the parameter $\beta$.}
\label{fig:LCAtr}
\end{figure}
\section{Numerical verification}
\label{sec:numvrf}
In the following section, we numerically verify the analytical results described in the former section. In Fig.~\ref{fig:LCAtr}(a), we are depicting the limit cycle attractor born in the switched system for $\beta = 0.01$, $A = 0.5$, $K_p = 1$, $K_d = 1$ and $\theta^* = 1$. In Fig.~\ref{fig:LCAtr}(b), we depict variation of the non-trivial Floquet multiplier corresponding to the limit cycle attractor as a function of parameter $\beta$. The non-trivial Floquet multiplier lies within the unit circle of the complex plane, and hence we get further numerical verification that the limit cycle born in the bifurcation is a stable orbit.

\begin{figure}
\centering
\includegraphics[scale=0.35]{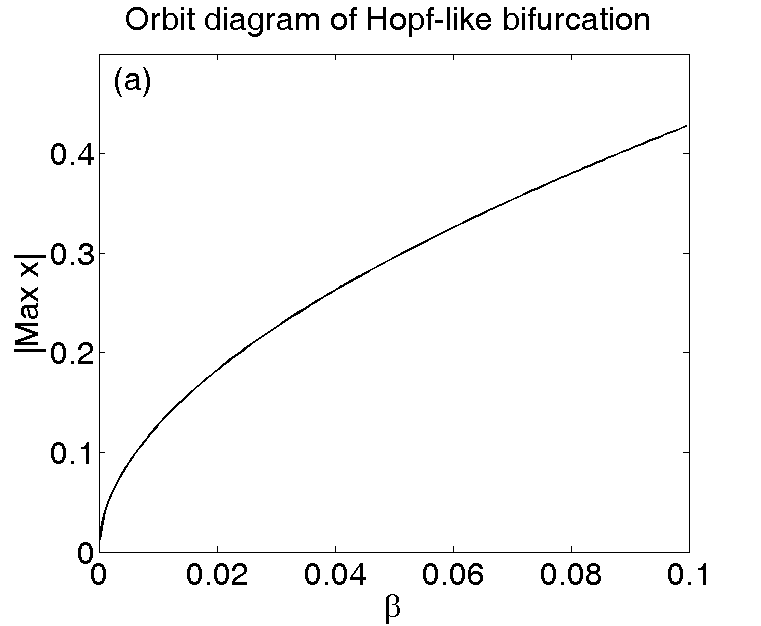}
\includegraphics[scale=0.35]{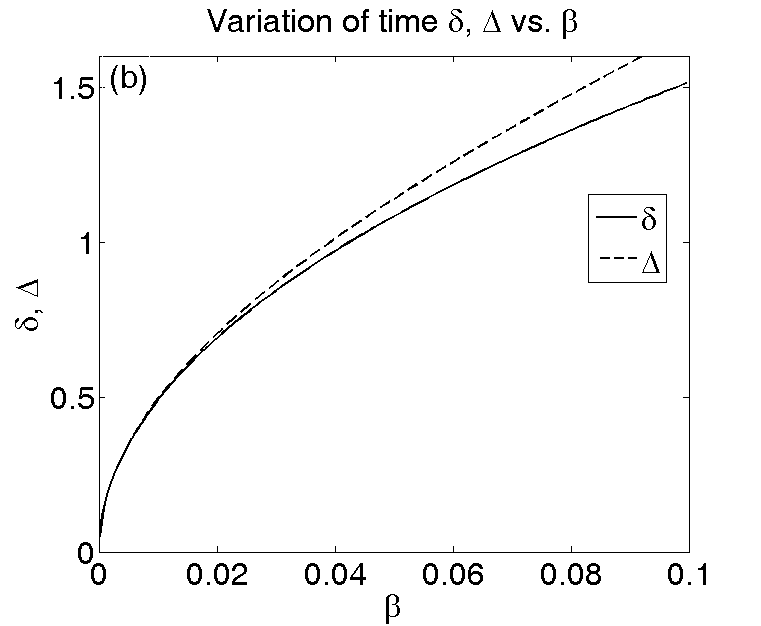}
\caption{(a) Bifurcation diagram depicting $|x_{\mbox{max}}|$ vs. $\beta$ on the limit cycle born in the Hopf bifurcation. (b) Time variation of the duration of one segment, generated by flow $\phi_1$, making up the limit cycle born in the Hopf bifurcation using numerical ($\Delta$) and asymptotic values ($\delta$ given by equation (\ref{eq:delT})).}
\label{fig:OD}
\end{figure}

 An orbit diagram where we depict
the variation of the maximum value of $\dot\theta$ on the limit cycle, say $|x_{\mbox{max}}|$, versus $\beta$ is then shown in Fig.~\ref{fig:OD}(a).

In Fig.~\ref{fig:OD}(a), we can clearly see the square root variation in the maximum value of the $\dot\theta = |x_{\mbox{max}}|$ component as a function of the bifurcation parameter $\beta$.
Then in Fig.~\ref{fig:OD}(b), the time variation required to generate a segment of the limit cycle by the flow $\psi_{OUT}$ as a function of parameter $\beta$ is compared with numerical values. Asymptotic convergence is clearly visible.
\section{Limit cycle attractor born due to delayed switching}
\label{sec:dlsys}
\subsection{System with delayed switching line and positional feedback}
In the context of human neuro-muscular control, there are always present neurological time delays. For example, in the context of human balance control, there is a time delay present
in the system due to neural processing and muscle activation. Therefore, we will compare
numerically the Hopf-like bifurcation that we analyzed for the planar case with the Hopf-like  scenario that we observed in switched system with dead-zone and time delay in the switching decision function. Namely consider a planar switched system of the
form
\begin{eqnarray}
\label{eq:1DL}
& & \ddot\theta - A \theta = 0,\quad |\theta(t-\tau)|\leq \theta^*,\\
\label{eq:2DL}
& & \ddot\theta - A \theta = -K_p\theta - K_d\dot\theta,\quad |\theta(t-\tau)| > \theta^*,
\end{eqnarray}
where $\theta^* > 0$, $K_p > A > 0$, $K_d > 0$, $K_p - A > K_d^2/4$ and $\tau = \Ord(\eps) > 0$.
That is we consider the same parameter space as in Sec.~\ref{sec:numvrf}. Similarly the flow $\psi_{IN}$ is
the solution of (\ref{eq:1DL}) and the flow $\psi_{OUT}$ is the solution of (\ref{eq:2DL}).

System (\ref{eq:1DL}) and (\ref{eq:2DL}) can be seen as a model for human balance control during quiet
standing where the neural transmission and muscle activation delays are included in the
delay of the switching decision function \cite{Asa:09,KoGletal:11}. The dead-zone  in the model can be seen as being related to the finite accuracy of sensing \cite{Asa:09,KoGletal:11}.
Define
\beq
\Sigma_+^{\psi_{IN}} = \{(\theta,\,\dot\theta)\in\RR^2 : \theta = \theta_1(\tau)\,\, \mbox{and}\,\,\dot\theta = \dot\theta_1(\tau) \},
\label{eq:dswlin}
\eeq
where
\begin{eqnarray*}
& & \theta_1(\tau) = \frac{\sqrt{A}\theta_0-\dot\theta_0}{2\sqrt{A}}\exp(-\sqrt{A}\tau) + \frac{\sqrt{A}\theta_0+\dot\theta_0}{2\sqrt{A}}\exp(\sqrt{A}\tau),
\\
& & \dot\theta_1(\tau) = -\frac{\sqrt{A}\theta_0-\dot\theta_0}{2}\exp(-\sqrt{A}\tau) + \frac{\sqrt{A}\theta_0+\dot\theta_0}{2}\exp(\sqrt{A}\tau),
\end{eqnarray*} are the images of the position and velocity states on $\Sigma_+$, namely for $\theta_0 = \theta^*$ and $\dot\theta_0\in\RR$, under the evolution of $\psi_{IN}$ for the small fixed time $\tau = \Ord(\eps)$.

Similarly, define
\beq
\Sigma_+^{\psi_{OUT}} = \{(\theta,\,\dot\theta)\in\RR^2 : \theta = \theta_2(\tau)\,\, \mbox{and}\,\,\dot\theta = \dot\theta_2(\tau) \},
\label{eq:dswlout}
\eeq
where
\begin{eqnarray*}
& & \theta_2(\tau) = \theta_0\exp(-a\tau)\cos(b\tau) + \frac{\dot\theta_0 + a\theta_0}{b}\exp(-a\tau)\sin(b\tau),\\
& & \dot\theta_2(\tau) = \dot\theta_0\exp(-a\tau)\cos(b\tau) - \frac{a\dot\theta_0 + a^2\theta_0 + b^2\theta_0}{b}\exp(-a\tau)\sin(b\tau),
\end{eqnarray*}
and $a^2 = K_d^2/4$, $b^2 = (K_p - A) - K_d^2/4$, $\theta_0 = \theta^*$ and $\dot\theta_0\in\RR$.
Thus $(\theta_2,\,\dot\theta_2)$ are the images of the position and velocity states, for any initial conditions on $\Sigma_+$, under the evolution of $\psi_{OUT}$ for the small fixed time $\tau = \Ord(\eps)$.
Since $\Sigma_+$ is a line and the flows $\psi_{IN}$ and $\psi_{OUT}$ are linear then $\Sigma_+^{\psi_{IN}}$ and $\Sigma_+^{\psi_{OUT}}$ are lines in the phase space $(\theta,\dot\theta)$ which, generically, will cross in some neighborhood of $(\theta^*,\, 0)$ for sufficiently small delay time $\tau$.
\subsection{Numerical observations}
\begin{figure}
\centering
 \includegraphics[scale=0.35]{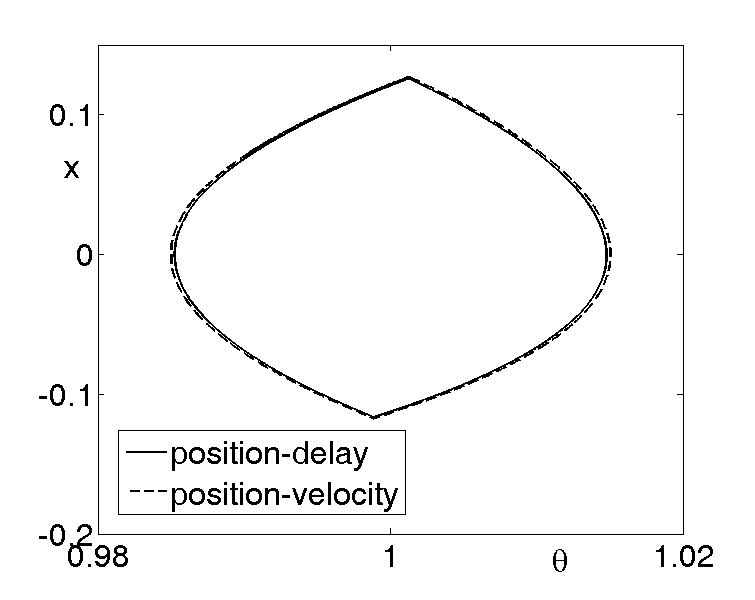}
 \caption{Limit cycle attractors born in Hopf-like bifurcation scenarios in switched (``position-velocity'') system (\ref{eq:sys1}) and (\ref{eq:sys2}) under the variation of the slope $\beta$ (dashed line) and in system (\ref{eq:1DL}), (\ref{eq:2DL}) with the delayed switching line (``position-delay'') under the variation of the delay time $\tau$ (solid line).}
 \label{fig:dsA}
 \end{figure}
We will investigate numerically a bifurcation in the model under the switching of the delay time $\tau$. For $\tau = 0$, we have a switched model with purely positional control and dead-zone, which has two stable pseudo-equilibria as the only attractors (see Sec.~\ref{sec:sysofint} and \cite{KoGletal:11}).
 For $\tau > 0$, we have a dynamical system which is infinite dimensional due to the fact that, to be able to determine the forward evolution, it is necessary to keep track of a segment of trajectory for $\theta$ state in the interval $[t - \tau,\, t]$.
 However, if certain genericity conditions are satisfied in our delayed switched model, the system dynamics reduces to the evolution of finite number of state variables (see \cite{Si:06, SiKoHodiB:10}). In particular, let $t_1$ be the time of evolution from any point
  $(\theta,\,\dot\theta)\in\Sigma_+^{\psi_{IN}}$,
 in some neighborhood of $(\theta^*,\, 0)$, to $\Sigma_+$, and $t_2$ be the time of evolution
from any point $(\theta,\,\dot\theta)\in\Sigma_+^{\psi_{OUT}}$ to $\Sigma_+$.

 Then for $\tau$ sufficiently small
 we may assume that $\tau < t_1$ and $\tau < t_2$. This implies that, assuming a limit cycle exists in some neighborhood of the point $(\theta^*,\, 0)$, the dynamics of the delayed switched system can be described locally by a smooth one-dimensional map $\Sigma_+^{\psi_{IN}} \mapsto \Sigma_+^{\psi_{IN}}$. Other Poincar\'e section, transversal to the limit cycle, can be chosen. What is important, however, is the reduction
of an infinite dimensional system to a finite dimensional one. In what follows, we show numerical evidence that this reduction preserves also quantitative features. In other words, we will provide numerical evidence that a switched system with position-velocity control may in certain circumstances behave exactly like a switched system with purely positional control and delayed switching.

\begin{figure}
\centering
\includegraphics[scale=0.35]{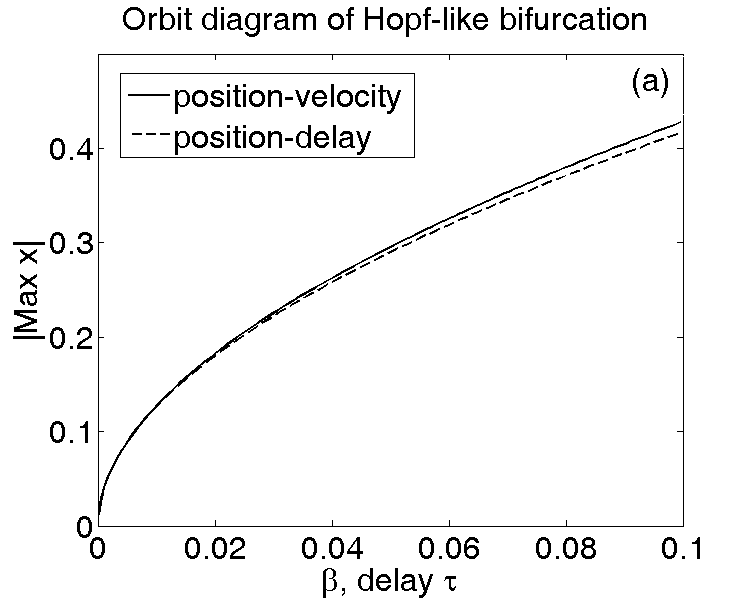}
\includegraphics[scale=0.35]{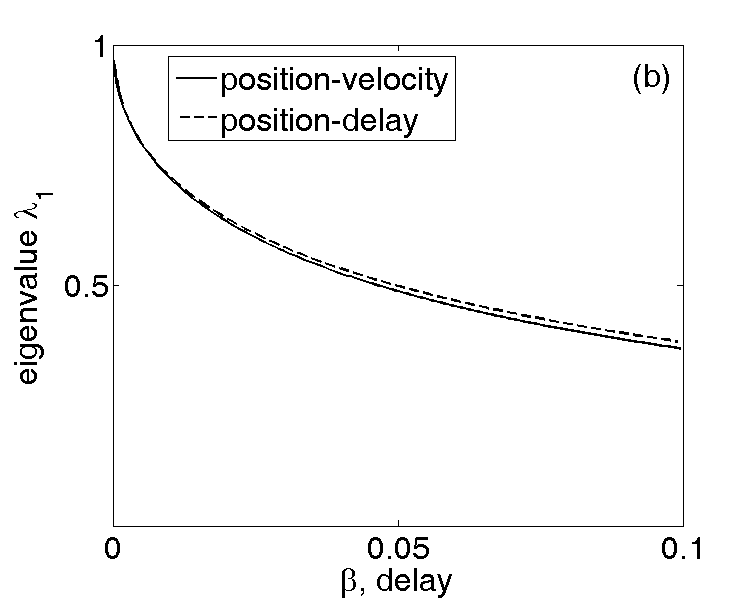}
\caption{(a)Comparison between the variation of the non-trivial Floquet multiplier of the limit cycle attractor in the system with delayed switching line (dashed line) and no delay (solid line).
(b) The variation of the non-trivial Floquet multiplier of the limit cycle attractor against the parameters $\beta$ and time delay}
\label{fig:dlfl}
\end{figure}

In Fig.~\ref{fig:dsA}, we
depict a limit cycle attractor born in the system for $\tau = 0.01$, $A = 0.5$, $K_p = 1$, $K_d =1$ and $\theta^* = 1$ and compare it to the limit cycle attractor born in the switched position-velocity system with $\beta = 0.01$. It can be seen that the two limit cycles are virtually indistinguishable. In Fig.\ref{fig:dlfl}(a) we compare bifurcation diagrams of both systems under variations of $\beta$ and $\tau$ from $0$, and in Fig.~\ref{fig:dlfl}(b) we verify the variation of the non-trivial Floquet multiplier born in the Hopf-like bifurcation in either system.
\section{Conclusions}
\label{sec:concl}
In the paper, we analyze a novel type of a discontinuity induced bifurcation in the case when
there is applied perturbation to the linear control feedback law. In particular, we show that switched linear systems with dead-zone and purely positional feedback under small parameter perturbation from position to position-velocity control may loose stable pseudo-equilibrium state (an equilibrium of a switched system which lies on the switching manifold) and produce limit cycle behavior in a Hopf-like scenario. Using asymptotic method we analyze this novel bifurcation and show the loss of stability of the pseudo-equilibrium and birth of stable limit cycles with the amplitude, say $|x|$, growing as a square root of the bifurcation parameter ($|x| = \Ord\sqrt{\beta}$, where $\beta$ refers to the bifurcation parameter). We should note that this bifurcation is different from the Hopf-like bifurcation scenario analyzed in \cite{FrPoRoTo:98}, where in a non-smooth Hopf-like bifurcation scenario the amplitude of the limit cycle born in the bifurcation grows linearly as a function of the bifurcation parameter. We verify our analysis numerically.

We then consider a switched system with dead-zone and purely positional feedback law, but with the switching decision function that contains a time delay. We investigate this system numerically for small values of the delay time ($\tau = \Ord(\eps)$). We find that the system, considering the same parameter values as the non-delayed one, exhibits a Hopf-like bifurcation scenario under the variation of delay time $\tau$, which non only qualitatively but also quantitatively matches the Hopf-bifurcation in the switched system with no time delay. In control literature, it has been suggested that delays in positional feedback laws may serve as approximation of velocity components since $v\approx (x(t+\tau) - x(t))/\tau$. However, in our case the time delay is included only in threshold detection and so the agreement of the qualitative and quantitative nature in the case of the two types of novel Hopf-bifurcation scenarios reported in the current work is somehow surprising. We should note here that similar systems have been analyzed in \cite{Asa:09,SiKuLi:12} and there an onset of small scale limit cycles born round pseudo-equilibria have been also reported. However, these systems have been characterized by the presence of time delays in the position and velocity state variables as well.

Future work is aimed at considering under what conditions the bifurcations analyzed in the current work are observed in higher dimensional switched systems. Also, the link between the Hopf-like bifurcation in position-velocity and position-delay switched systems, which we report numerically in the current work, will be analyzed.
From the application point of view, we are interested in investigating whether there is a link between the small scale stable oscillations born due to a small change in the character of the control law in switched models with dead-zone and the dynamics of neuromotorcontrol systems. In particular, we are interested in understanding sway motion during quiet standing of humans affected by diabetes by linking changes of sway patterns with bifurcations analyzed in the current work.


\end{document}